\documentclass[twoside,a4paper]{article}

\usepackage{cite}
\usepackage{amsmath}
\usepackage{amssymb}
\usepackage{graphicx}
\usepackage{psfrag}
\usepackage[T2A]{fontenc}
\usepackage[cp1251]{inputenc}

\usepackage{textcomp}

\usepackage{color}

\usepackage[eqsecnum]{cmpj2}

\newcommand{\ev}{\mbox{{eV}}}
\newcommand{\mev}{\mbox{meV}}
\newcommand{\oss}{\mbox{O$_{s}$}}
\newcommand{\osp}{\mbox{O$_{s}^+$}}
\newcommand{\osm}{\mbox{O$_{s}^-$}}
\newcommand{\osn}{\mbox{O$_{s}^0$}}
\newcommand{\nss}{\mbox{N$_{s}$}}
\newcommand{\mhz}{\mbox{MHz}}

\issue{2016}{19}{3}{33301}
\doinumber{10.5488/CMP.19.33301}

\title[DFT study of substitutional oxygen in diamond]%
{Density functional theory study of substitutional oxygen in diamond}%

\author[K.M.~Etmimi \textsl{et al.}]{K.M.~Etmimi\refaddr{label1},
        P.R.~Briddon\refaddr{label2}, A.M.~Abutruma\refaddr{label3}, A.~Sghayer\refaddr{label1}, S.S.~Farhat\refaddr{label1}}
\addresses{
\addr{label1} Physics Department, Faculty of Science, University of Tripoli, Tripoli, Libya
\addr{label2} School of Electrical, Electronic and Computer Engineering
  Newcastle University, \\ Newcastle upon Tyne, England, NE1~7RU, UK
\addr{label3} High Institute for Comprehensive Professions (Al-Shmokh Institution), Tripoli, Libya
}

\authorcopyright{K.M.~Etmimi, P.R.~Briddon, A.M.~Abutruma, A.~Sghayer, S.S.~Farhat, 2016}
\date{Received July 20, 2015, in final form December 22, 2015}

\begin{document}

\maketitle

\begin{abstract}
A few studies have been recently presented for the existence of  oxygen in diamond,
for example, the N3 EPR centres have been theoretically and experimentally assigned the model made up
 from complex of substitutional nitrogen  and substitutional oxygen as nearest neighbours.
We present ab initio calculations of substitutional oxygen in diamond
in terms of stability, electronic structures, geometry and hyperfine interaction
and show that substitutional oxygen with C$_{2v}\,$, $S=1$
is the ground state configuration.
We find that oxygen produces either a donor or acceptor level depending on the position of the Fermi level.
\pacs  31.15.E-,  81.05.ug, 31.30.Gs
\end{abstract}

\section{Introduction}

High electron and hole mobilities at
room temperature and unrivalled thermal conductivity at
room temperature, mean that diamond could be the material
of choice for high-power and high-frequency electronics.
Many defects in a diamond such as nitrogen~\cite{briddon-PB-185-179,jones-DRM-53-35,atumi-JPCM-25-6-065802} and boron~\cite{chrenko-PRB.7.4560} are the main chemical elements
now well identified. Nitrogen as a simple
substitutional defect forms a deep level at 1.7~\ev~\cite{farrer-SSC-7-685} below the conduction band edge and
the boron gives a shallower level at 0.37~{\ev} above the edge of the valence band~\cite{crowther-PR-154-772}.
For this reason, a variety of other chemical
defects in a diamond are now being investigated.

Oxygen is expected to be one of important impurities in a diamond
due to its relatively close size of carbon atom and abundance,
and its has been suggested to lead to n-type conductivity in a diamond~\cite{prins-PRB-61-7191}.
Experimentally, oxygen was found in mineral inclusions in a diamond~\cite{Walker-N-263-275}
and combustion analysis of converting diamond to graphite indicated
 high levels of oxygen in natural diamond~\cite{melton-n-263-309}.
Moreover, a small number of optical centres
may be related to oxygen~\cite{ruan-APL-60-1379,mori-apl-60-47}.

Electron-paramagnetic-resonance (EPR) spectroscopy is a powerful probe used
to identify enormous  centres in diamond with nucleus spin of none zero.
A small percentage (0.04$\%$)
of natural oxygen is~$^{17}$O \cite{iakoubovskii-PRB-66-045406} which is
consistent with little evidence of the
involvement of oxygen in electron paramagnetic resonance centres in diamond~\cite{Ammerlaan1998}.
However, enrichment of diamond in the $^{17}$O isotope can occur
during diamond growth~\cite{iakoubovskii-PRB-66-045406} or $^{17}$O ion implantation~\cite{prins-PRB-61-7191,iakoubovskii-PRB-66-045406}
where the former undergoes insufficient control of the gas environment during diamond growth~\cite{iakoubovskii-PRB-66-045406},
and the latter is induced by lattice damage where the oxygen may be trapped for instant vacancy.
EPR centre called KUL12 was detected to be an $S=1/2$ centre interacting with one $I=5/2$ nucleus
with $A_\parallel = - 362$~MHz and $A_\perp = - 315$~MHz.
Unfortunately, there is no other direct measurement of this centre.

The N3 and OK1 centres have been suggested to contain oxygen; the
former in a nearest neighbour nitrogen-oxygen pair, and the latter
in a second nearest neighbour nitrogen-oxygen pair.
Previously, we analysed N3 and OK1 in a broader study~\cite{etmimi-jpcm-22-385502}
and we concluded that the most suitable candidate structure for N3 is N$_s$--O$_s$.
For OK1, none of the proposed models yield hyperfine
tensors in agreement with experiment.
Three of the EPR centres found in synthetic diamonds grown in carbonate medium
 in recent experimental study~\cite{komarovskikh-PSSA-210-2074}, using the high pressure
apparatus BARS~\cite{palyanov-7-3169,Komarovskikh-PSSA-31163}
showed that OX1, OX2 and OX3 centres are oxygen atoms occupying substitutional, interstitial and next to vacancy sites, respectively.

The previous theoretical work using ab initio calculations~\cite{gali-JPCM-13-11607} shows that substitutional
 oxygen exhibits carbon vacancy character which gives rise to an occupied $a$-level into
the middle of the band gap and unoccupied $t$-level just below the conduction band.
A theoretical work has predicted that oxygen introduces a mid-gap
donor level in the band gap of a diamond, which is above the
fundamental level of vacancy being 2~{\ev} above the top of valence band.
So, in a material containing both types of centres, one would expect a charge transfer to occur,
which gives rise to EPR active defect.

In this work, extensive calculations on  different models containing oxygen atoms were carried out, and the total energies and other
properties of defects were determined using ab initio calculations.

\section{Method}

The structures were modelled using density-functional
calculations with the exchange-corre\-la\-tion in a generalised gradient
approximation~\cite{perdew-prl-77-3865} by the AIMPRO
code~\cite{briddon-pssb-217-131,rayson-cpc-178-128}.  The
Brillouin-zone is sampled using the Monkhorst-Pack
scheme~\cite{monkhorst-prb-13-5188} with a uniform mesh of
 $2\times2\times2$ special $k$-points.
For several sample structures, we calculated the total
energies using a $4\times4\times4$ mesh, which indicated that the
relative total energies are converged to better than 10~\mev.

The valence states  were represented by a  set of  atom-centred $s$- and $p$- with the addition of a
set of $d$-like Gaussian functions~\cite{goss-tap-104-69} to allow for polarization,
and the Kohn-Sham states were expanded  with the help of a contracted basis with a total of 22~functions on each carbon and oxygen atom.
For the charge density evaluation, the plane waves with a
cut-off of 300~Ha were used, yielding structures optimized until the total energy changes by less than
$10^{-5}$~Ha.
The lattice constant and the bulk modulus were within ~1\%
and 2\%, respectively, of experimentally determined values. The lattice constant was optimized, keeping the symmetry of the supercell fixed,
giving a value of 3.5719~\AA, close to the experimental value of 3.5667~\AA~\cite{Sze1981}.
The calculated direct and indirect band gaps agree with the
published plane-wave values~\cite{liberman-PRB-62-6851} (5.68 and 4.18~\ev).

In general 216-atom, simple-cubic supercells of side length $3a_0$ are used.
Core-electrons are eliminated by using pseudo-potentials~\cite{hartwigsen-prb-58-3641}, the $1s$ electrons of C and O are in the core, and the
$3p$ electrons are treated as the valence ones, so that hyperfine
interactions are obtained by reconstructing the all-electron wave
functions in the core region~\cite{shaw-prl-95-105502,blochl-prb-50-17953}.  The atomic
calculations for the reconstruction in the hyperfine calculations were
performed using a systematic polynomial
basis~\cite{rayson-pre-76-026704}.
Electrical levels were calculated using the marker method by comparing acceptor and donor
with B and N, respectively.

\section{Results and dissections}

\begin{figure}[!t]
\psfrag{a}{1.45}
\psfrag{b}{2.02}
\psfrag{c}{1.72}
\psfrag{d}{1.74}
\psfrag{e}{1.73}
\psfrag{f}{1.64}
\psfrag{g}{1.82}
\psfrag{h}{2.08}
\psfrag{i}{1.63}
\psfrag{j}{1.74}
\psfrag{o}{O}
\psfrag{c1}{C1}
\psfrag{c2}{C2}
\psfrag{c3}{C3}
\psfrag{c4}{C4}
\center
  \begin{minipage}{0.41\textwidth}
   \centerline{(a)}
    \includegraphics[width=\textwidth,clip]{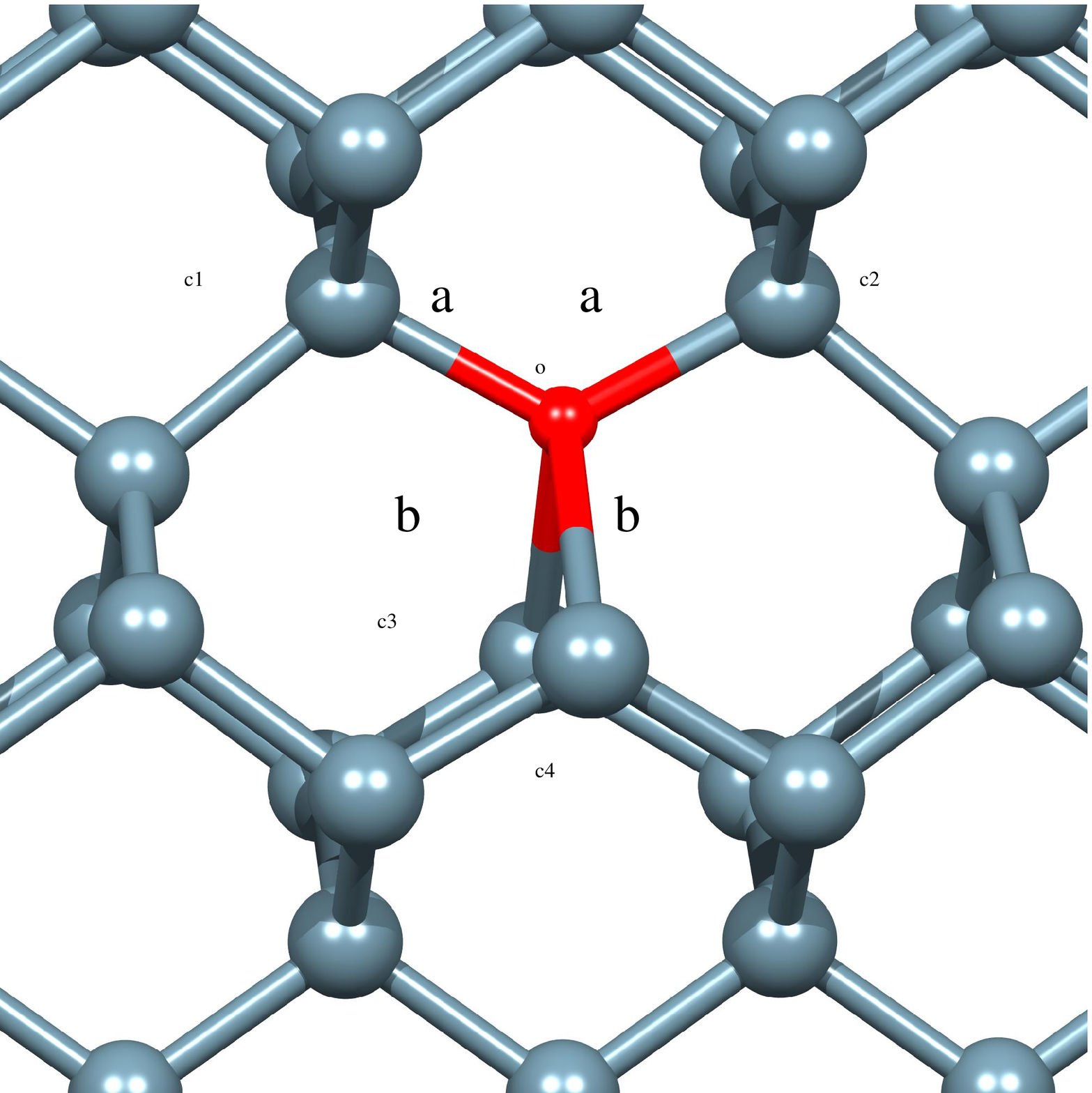}
  \end{minipage}
\begin{minipage}{0.41\textwidth}
   \centerline{(b)}
    \includegraphics[width=\textwidth,clip]{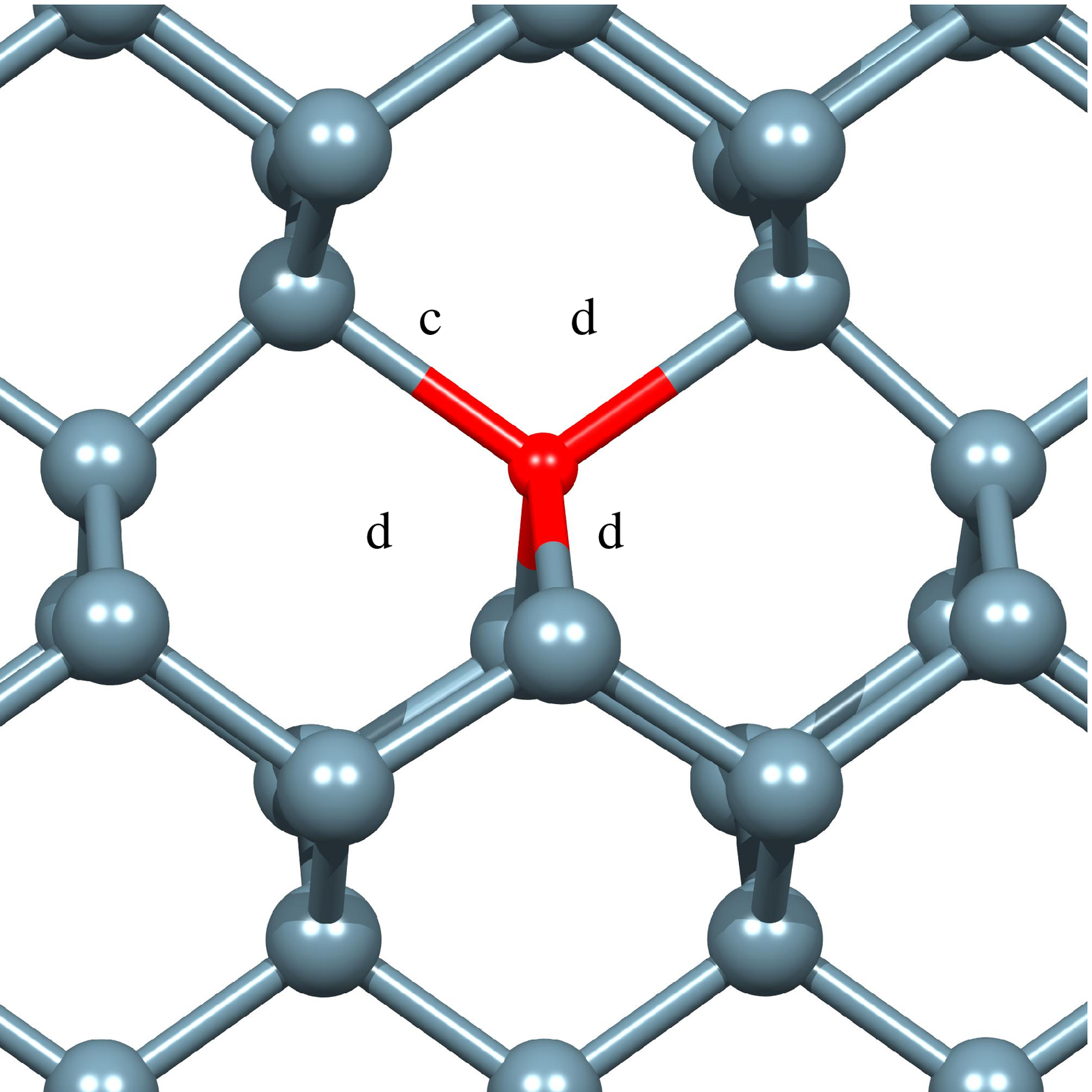}
  \end{minipage}
\begin{minipage}{0.41\textwidth}
  \centerline{(c)}
    \includegraphics[width=\textwidth,clip]{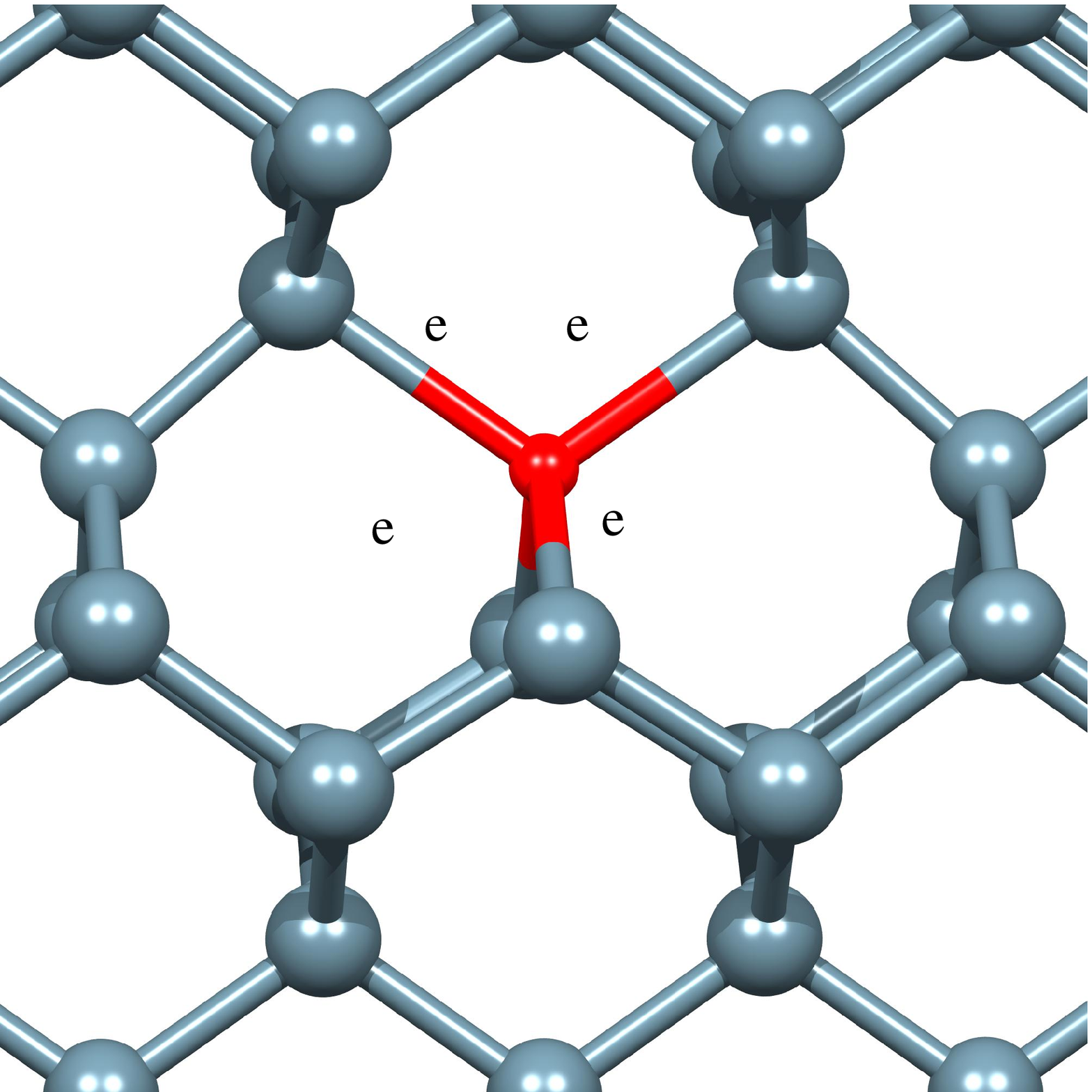}
  \end{minipage}
\begin{minipage}{0.41\textwidth}
   \centerline{(d)}
    \includegraphics[width=\textwidth,clip]{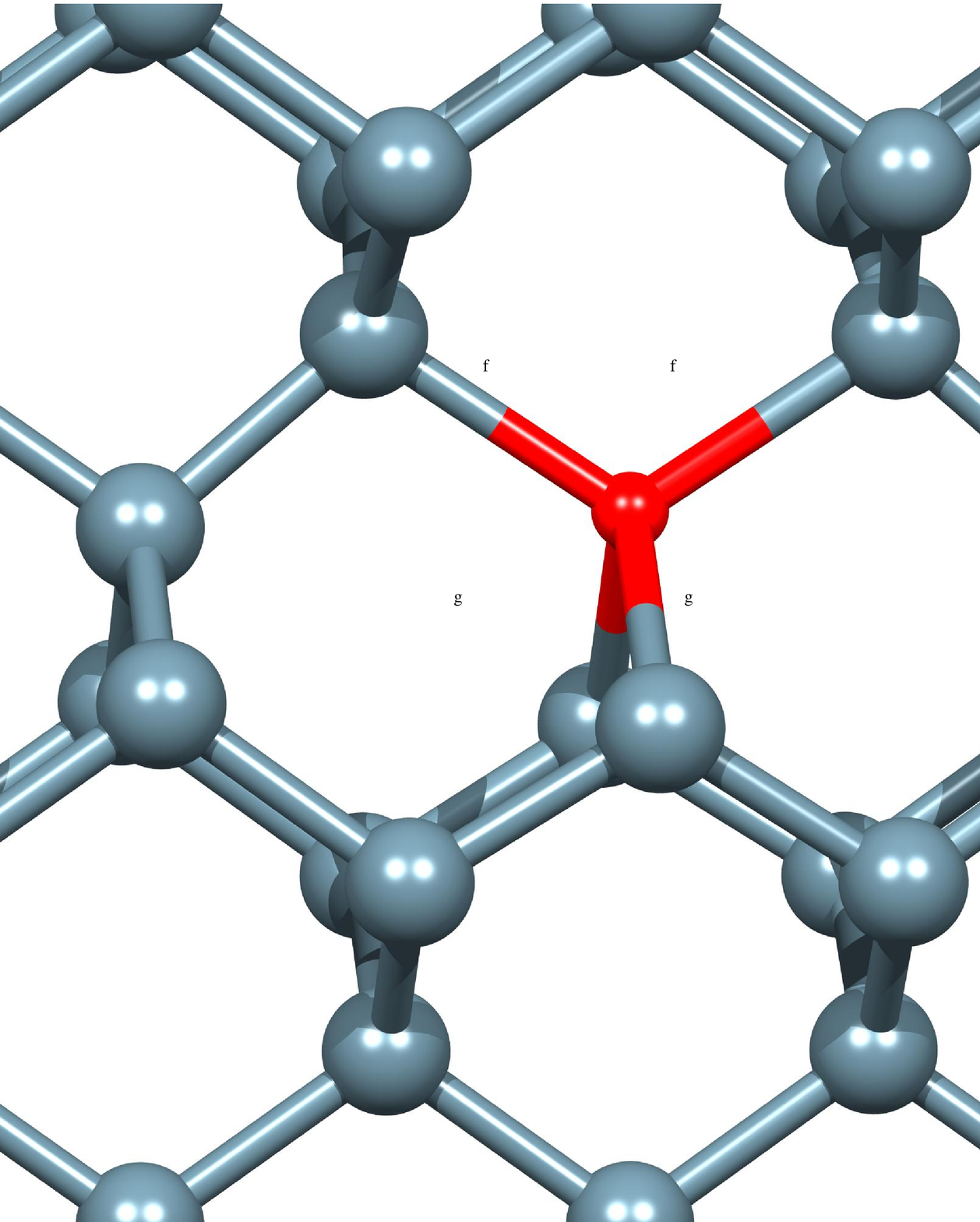}
  \end{minipage}
\begin{minipage}{0.41\textwidth}
   \centerline{(e)}
    \includegraphics[width=\textwidth,clip]{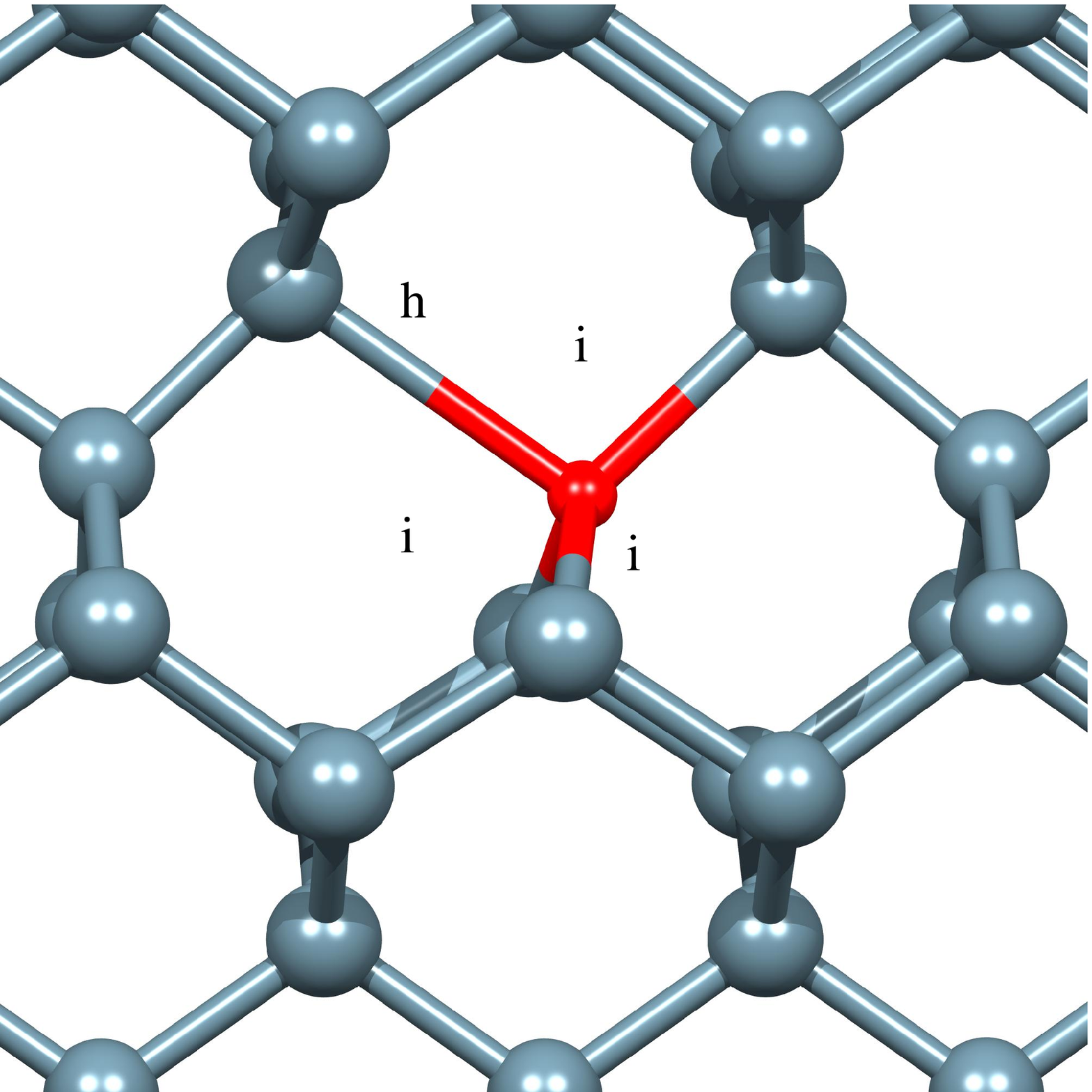}
  \end{minipage}
  \caption{ (Color online) Schematic structures of the substitutional oxygen in both the $S=0$ and $S=1$ configurations.
 Grey and red spheres represent C and O. (a) $C_{2v}$, $S=1$, (b) $C_{3v}$, $S=0$, (c) $T_d$, $S=0$, (d) $C_{2v}$, $S=0$,
(e) $C_{3v}$, $S=1$.
 Bond lengths in \AA.}
\label{fig:O}
\end{figure}

Different charged forms {\osn}, {\osp} and {\osm} of substitutional oxygen in diamond are examined.
In all cases, the $C_{2v}$ configuration is found to be
favoured.

In the neutral charge state, we find several metastable structures for substitutional O$^0_s$ in a diamond.
Interestingly, the spin orientation was crucial in terms of
determining the stability.
The lowest in energy exhibits a $C_{2v}$ $S=1$, as schematically shown in
 figure~\ref{fig:O}~(a). The oxygen atom undergoes
a distortion along $\langle$001$\rangle$,
the oxygen moves strongly off centre to form two C--O bonds leaving behind two C dangling orbitals.
This suggests that it may undergo a symmetry lowering distortion, probably of a chemical re-bonding.
Three structures [figures~\ref{fig:O}~(b), (c) and (d)] were found to be energetically indistinguishable and higher
in energy than the ground state configuration by just 0.2~{\ev}. The oxygen atom in figure~\ref{fig:O}~(c) is in a
fourfold coordinated arrangement with C--O bonds of lengths 1.73~{\AA}.

A trigonally symmetric state was also examined and was found to be metastable.
The $C_{3v}$ $S=0$ structure in the figure~\ref{fig:O}~(b) exhibits a slight displacement along $\langle$111$\rangle$,
where one of the four neighbouring C~atoms move from 1.73~{\AA} to 1.72~{\AA} compared
to the on-site structure in figure~\ref{fig:O}~(c).
The energy differences [between figure~\ref{fig:O}~(b) and (c)] are just of
 a few {\mev}.
Symmetrically equivalent to the ground state configuration [figure~\ref{fig:O}~(a)], the spin averaged structure with two equivalent
carbon neighbours to O is moved closer to the impurity, lowering the symmetry to $C_{2v}$,
however, it is 0.21~{\ev} higher in energy than the most stable one.

The structure in the figure~\ref{fig:O}~(e) exhibits $C_{3v}$ $S=1$ where the oxygen is significantly
displaced off-site along $\langle$111$\rangle$ producing an elongated C--O bond
to 2.08~{\AA} which significantly increased the energy to 0.57~{\ev} compared to the most stable configuration.

Generally, the substitutional O atom bonds relatively weakly to the carbon dangling bonds in the vacancy,
the reason for this probably being the oxygen atom having a relatively small atom compared to carbon
and it can be understood as having vacancy-like characteristics,  which has proved successful in explaining the
electronic structure of the defects in a diamond~\cite{watkins-PBC-117-9}.
\begin{table}[!t]
  \caption{Relative energies for different structures with different
 symmetries of O$_s$.
 The zero of energy for neutral, positive and negative charged state is set
to the O$_s^0$ $S=1$ $C_{2v}$, O$_s^{+1}$ $S=1/2$ $C_{2v}$ and O$_s^{-1}$
$S=1/2$ $C_{2v}$,
respectively.}
\label{table:OHFI}
\vspace{2ex}
\center
  \begin{tabular}{|c|c|c|c|}
   \hline
\hline
    Symmetry&Charge state&Spin configuration&Relative energy  \\
\hline
\hline
$C_{3v}$ &neutral&$S=0$&$0.20$~{\ev}\\
\hline
$T_d$&neutral&$S=0$&$0.20$~{\ev}\\
\hline
$C_{2v}$&neutral&$S=0$&$0.21$~{\ev}\\
\hline
$C_{3v}$&neutral&$S=1$&$0.57$~{\ev}\\
\hline
$C_{3v}$&positive&$S=1/2$&0.04~{\ev}\\
\hline
$T_d$&positive&$S=1/2$&0.05~{\ev}\\
\hline
$C_{3v}$&negative&$S=1/2$&0.57~{\ev}\\
\hline
$T_d$&negative&$S=1/2$&1.13~{\ev}\\

   \hline
\hline
  \end{tabular}
\end{table}

In the positive charge state, three different symmetry configurations are obtained.
The differences are within a few {\mev} as listed in table~\ref{table:OHFI}. The
lowest structure has $C_{2v}$ symmetry lower than two other structures by just
0.04~{\ev} for $C_{3v}$ and 0.05~{\ev} for $T_d$ configuration, respectively, where all structures with spin $S=1/2$.
 The energy difference is so tiny that one cannot be certain which possesses the
lowest energy.
Similarly to {\osp}, the $C_{2v}$ $S = 1/2$ configuration is found to be
favoured in the negatively charged state.
Moreover, there are two other metastable configurations which are high in energy
as listed in table~\ref{table:OHFI}.

\begin{figure}[!b]
\psfrag{a}{$a_1^*$}
\psfrag{t}{$t_2^*$}
\psfrag{a1}{$a_1$}
\psfrag{b1}{$b_1$}
\psfrag{b2}{$b_2$}
\vspace{-2mm}
\center
  \begin{minipage}{0.49\textwidth}
   \centerline{(a)}
    \includegraphics[width=\textwidth,clip]{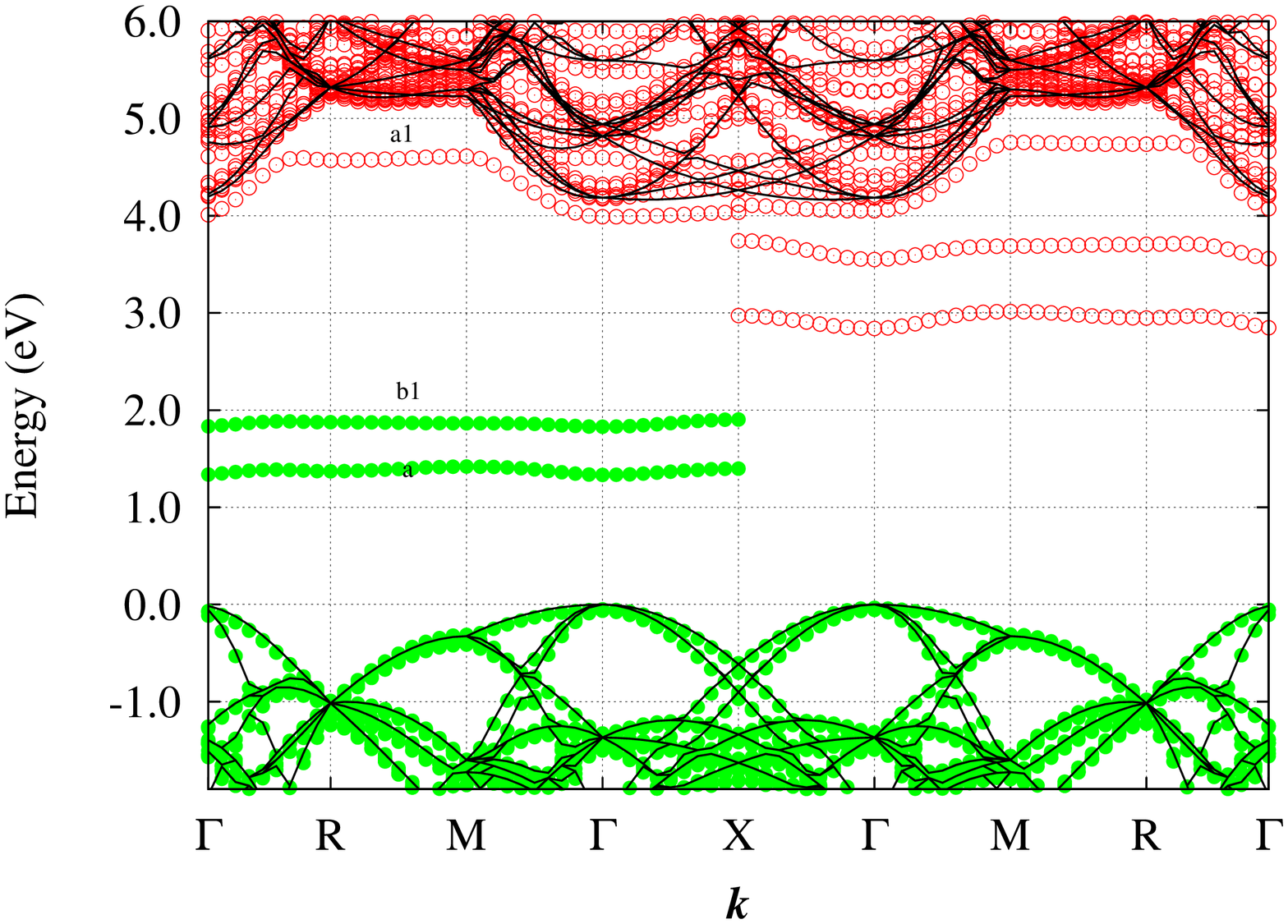}
  \end{minipage}
\begin{minipage}{0.49\textwidth}
   \centerline{(b)}
    \includegraphics[width=\textwidth,clip]{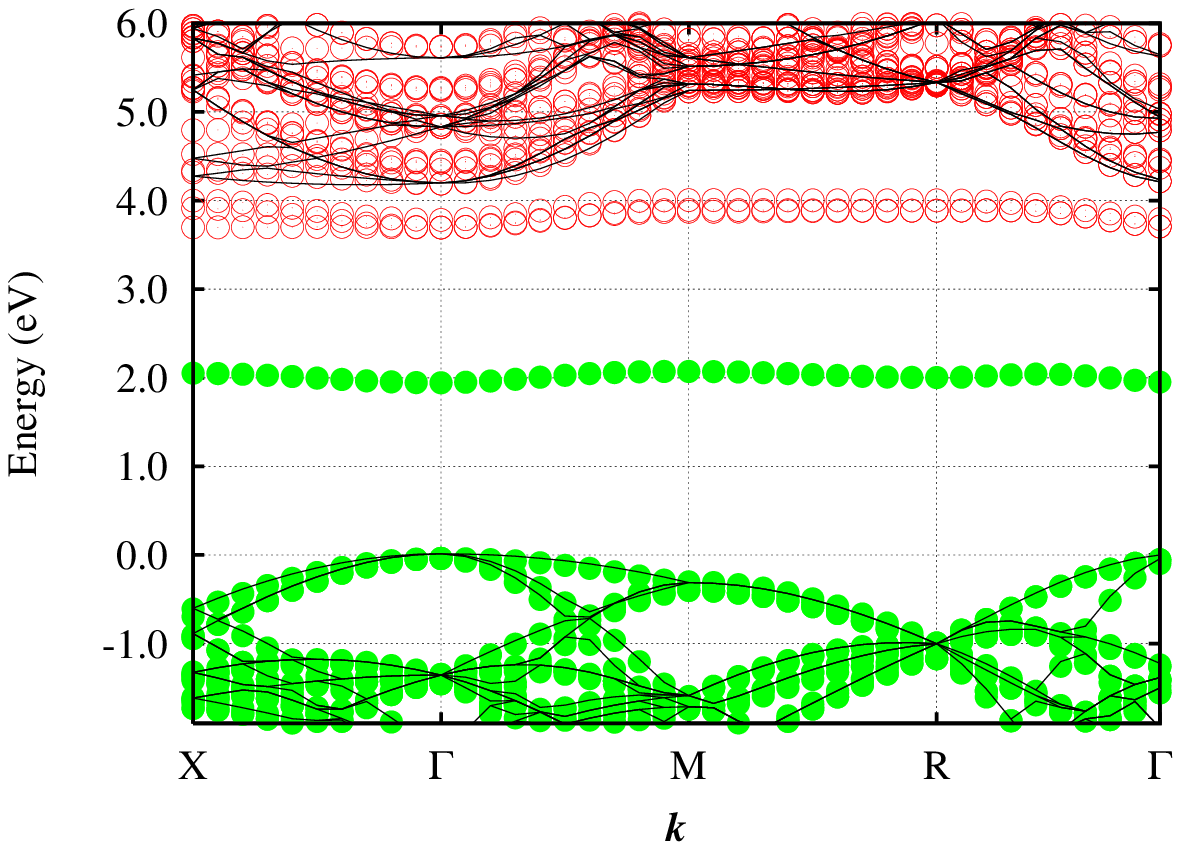}
  \end{minipage}
\begin{minipage}{0.49\textwidth}
   \centerline{(c)}
    \includegraphics[width=\textwidth,clip]{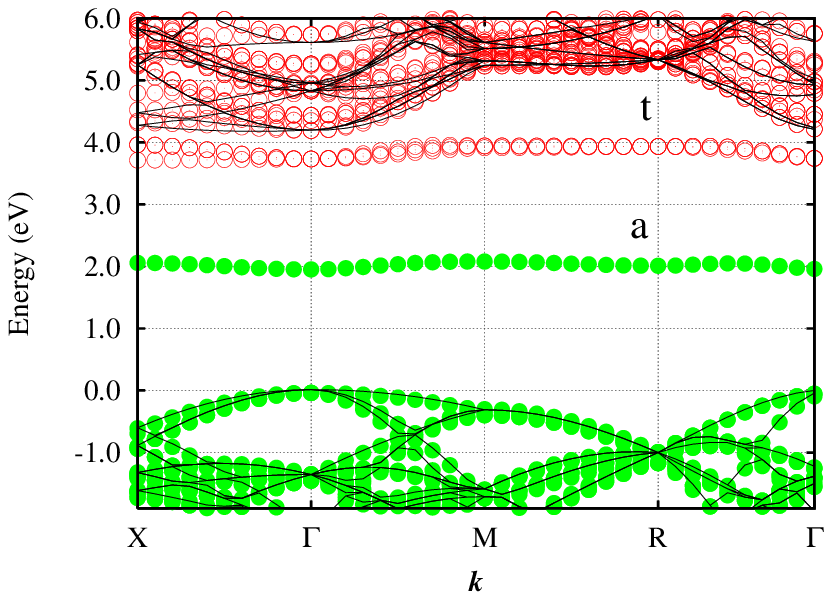}
  \end{minipage}
  \caption{ (Color online) Band structure for $S=1$ and $S=0$
  configurations of substitutional oxygen in diamond.  Filled and empty
circles show filled and empty bands, respectively.
 The energy scale is defined by the valence band top
at zero energy. (a) $C_{2v}$ $S=1$, (b) $C_{3v}$ $S=0$,
 (c) $T_d$ $S=0$.}
   \label{fig:bandest-O}
\end{figure}

Electronically, we present the previously calculated energy structure of on-site configuration
quoted by Gali et~al.~\cite{gali-JPCM-13-11607} that possesses the ground state configuration
[figure~\ref{fig:bandest-O}~(c)]. Its band gap consists of
 one fully  non-degenerate one-electron states $a_1$ near the middle of the band gap and
one triplet degenerate~$t_2$ state close to the conduction band, with a total
occupation of two electrons, where $a_1$ and $t_2$
 levels are introduced in the band gap due to C vacancy~\cite{coulson-PRSLA-241-433}.
Apparently, the $a_1$ and $t_2$ levels hybridized with oxygen valence states $s$ and $p$ atomic orbitals,
which gives rise to four levels $a_1$, $t_2$, $a^*_1$ and $t^*_2$, two states will be
in the band gap, the anti-bonding $a^*_1$ level will be occupied by two electrons
and triplet degeneracy $t^*_2$ will be unoccupied.

Our calculation shows that $C_{2v}$ $S=1$ in the neutral charge state is the most stable structure,
where the $t_2$-level
 is split into three levels, $b_1a_1b_2$ as shown schematically in
figure~\ref{fig:bandest-O}~(a),  where $b_2$ lies in the conduction band.
 The $b_1$ and $a_1$ levels are odd and even combinations of the neighbouring carbon
dangling bonds which are farther apart than the other two to oxygen, respectively.

Other previous density functional calculations~\cite{gali-JPCM-13-11607} found that O$_s$ has $T_d$ symmetry and is the most stable
configuration in neutral, negative and
positive charge states. However, it is unclear whether the $S=1$ was considered.
We find that the spin states for the neutral and positively charged states follow the Hund rule,
so the positive and neutral defects have $S = 1/2$ and $S = 1$, respectively.
Compared to the band gap of the well-known substitutional nitrogen (it has $C_{3v}$), O$_s$ with $C_{3v}$
$S=0$ symmetry has more levels. In addition to $a_1$ level associated with the radical on the unique
 carbon, there is one empty doublet degeneracy $e^0$ and
 a singlet empty $a^0$ from anti-bonding between O and there are three identical carbon
 atoms neighbouring the oxygen as shown in figure~\ref{fig:bandest-O}~(b).
The oxygen atom is more shared
with the three close neighbours.

{\oss} is theoretically electrically active, with donor, double
donor and acceptor
levels being estimated to be at $E_c-2.8~{\ev}$, $E_v+0.04~{\ev}$ and $E_c-1.9~{\ev}$, respectively. These
levels are in disagreement with the previous density functional
calculation~\cite{gali-JPCM-13-11607} of values $E_v+1.97~{\ev}$,
$E_v+1.39~{\ev}$ and $E_v+2.89~{\ev}$, respectively.
Oxygen can exhibit an amphoteric behaviour depending on the location of Fermi level.
Since the acceptor level of oxygen lies below a donor level such as {\nss} and the donor level
of oxygen lies above the acceptor level such as  $V$~\cite{dannefaer-DRM-10-2113}, one would
 expect the charge transfer to occur in the material containing both types of centres.
Previously, in~\cite{etmimi-jpcm-22-385502} we showed that Nitrogen-Oxygen complexes render both an acceptor and donor; we find the $(-/0)$ and $(0/+)$
levels at $E_v+ 3.7~{\ev}$ and $E_v + 1.5~{\ev}$.
\begin{table}[!b]
  \caption{Calculated hyperfine tensors (MHz) of  oxygen and
    the four nearest neighbour $^{13}$C in substitutional O. $\theta$ and $\phi$
are given relative to the directions $\langle0 0 1\rangle$ and $\langle1 0 0\rangle$, respectively.}
\label{table:O2HFI}
\vspace{2ex}
\center
  \begin{tabular}{|c|c|c|c|c|c|c|}
 \hline \hline
Species
    & \multicolumn{2}{c|}{$A_1$}
    & \multicolumn{2}{c|}{$A_2$}
    & \multicolumn{2}{c|}{$A_3$}\\    \hline \hline
\multicolumn{7}{|c|}{$C_{2v}$, $S=1$}\\
\hline
$^{17}$O            &$-198$ &$(90,315)$  &$-181$  &$(00,00)$  &$-165$&$(90,45)$\\ \hline
C1           &$-1$ &$(147,225)$&$-1$  &$(90,135)$  &$4$   &$(123,45)$\\ \hline
C2           &$-1$ &$(147,45)$  &$-1$  &$(90,135)$  &$4$   &$(57,45)$\\ \hline
C3           &$205$ &$(126,315)$ &$86$   &$(90,45)$  &$86$    &$(144,135)$\\ \hline
C4           &$205$ &$(54,315)$&$86$   &$(90,45)$  &$86$    &$(144,315)$\\
\hline
\multicolumn{7}{|c|}{$C_{2v}$, charge$=+1$, $S=1/2$}\\\hline
$^{17}$O            &$-568$ &$(00,129)$ &$-552$  &$(90,45)$ &$-538$ &$(90,315)$\\\hline
C1           &$41$   &$(125,45)$ &$18$    &$(102,144)$  &$18$  &$(142,250)$\\\hline
C2           &$41$   &$(55,45)$ &$18$    &$(102,126)$  &$18$   &$(142,20)$\\\hline
C3           &$197$   &$(125,315)$ &$86$    &$(90,45)$  &$86$   &$(145,135)$\\\hline
C4           &$197$   &$(55,315)$ &$86$    &$(90,45)$  &$86$   &$(145,315)$\\

\hline
\multicolumn{7}{|c|}{$C_{2v}$, charge$=-1$, $S=1/2$}\\\hline
$^{17}$O            &$-19$ &$(90,135)$ &$31$  &$(00,00)$ &$40$ &$(90,45)$\\\hline
C1           &$-7$   &$(82,225)$ &$-4$    &$(172,224)$  &$-4$  &$(90,315)$\\\hline
C2           &$-7$   &$(98,225)$ &$-4$    &$(172,46)$  &$-4$   &$(90,315)$\\\hline
C3           &$238$   &$(126,315)$ &$105$    &$(90,45)$  &$105$   &$(144,135)$\\\hline
C4           &$238$   &$(54,315)$ &$105$    &$(90,45)$  &$105$   &$(144,315)$\\
 \hline \hline
  \end{tabular}
\end{table}

\begin{figure}[!t]
\center
  \begin{minipage}{0.40\textwidth}
    \includegraphics[width=\textwidth,clip]{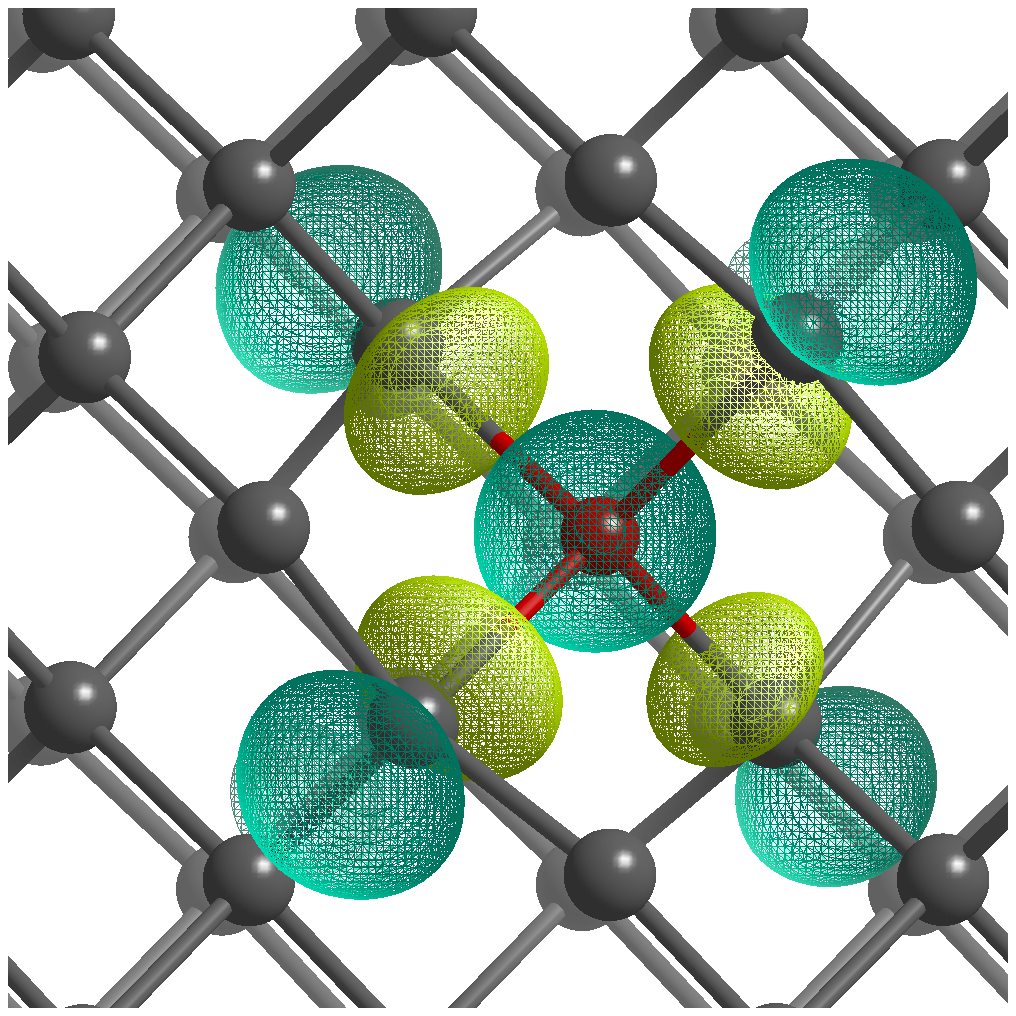}
  \end{minipage}
  \caption{ (Color online) Unpaired electron Kohn-Sham functions for positive substitutional oxygen with C$_{3v}$ symmetry.}
   \label{fig:O-positive-C3v}
\end{figure}

Hyperfine tensors for the most stable configurations within different charge states of oxygen and four
nearest neighbours are listed in table~\ref{table:O2HFI}.
There are no hyperfine values for substitutional oxygen in literature so far.
{\osp} with $C_{3v}$ structurally resembles P1 centre, although the present
 calculation shows that this configuration is metastable within a few~{\mev}.
 $C_{3v}$ symmetry means that one of the four neighbouring C~atoms is farther away from the oxygen than the other three,
the hyperfine tensor on these carbon atoms is small ($A_\parallel=89~\mhz$, ${A_\perp=42~\mhz}$) compared to those in P1 centre, which means
that the spin density is not mostly localized on the carbon radical site. It is distributed
 on the anti-bond between O and four neighbouring carbon atoms as shown in figure~\ref{fig:O-positive-C3v}.
In the negatively charged state with $C_{2v}$ symmetry, the spin density is
strongly localized in the vicinity of the carbon radical sites, leading to
small, anisotropic hyperfine tensors for the oxygen,
whereas in neutral charged state, the relatively larger values for
the hyperfine O compared to the negatively charged state are due to the relatively big
amount of spin density on the
O-site, and to some extent on the carbon radical atoms, from odd and even
 combinations of the neighbouring carbon dangling bonds for the highest and second
 highest occupied levels, respectively.
In the positively charged state, the even
 combinations of neighbouring carbon dangling bonds make the values of hyperfine tensor
 on O still larger.

\section{Conclusions}
We have used ab initio computational modelling mainly for the stability
and electronic structure on different forms of the substitutional oxygen in diamond.
Energetically we find that $S=0$ $C_{2v}$ is the most stable structure where both $S=0$ and $S=1$ are considered.

The band gap of substitutional oxygen gives rise  to two states, one $a_1$ state located near the middle of the band gap
and the other $t_2$ state located close to the conduction band edge. The $t_2$ state is populated when O becomes negatively charged
or neutral $S=1$ configuration.

\ukrainianpart
\title{Дослідження методом теорії функціоналу густини кисню заміщення  в алмазі}
\author{K.M.~Етмімі\refaddr{label1},
        П.Р. Бріддон\refaddr{label2}, A.M.~Абутрума\refaddr{label3}, A.~Сайер\refaddr{label1}, С.С.~Фархат\refaddr{label1}}
\addresses{
\addr{label1} Фізичний відділ, факультет природничих наук, Університет Тріполі, Тріполі, Лівія
\addr{label2} Коледж електричної, електронної і комп'ютерної інженерії, Університет Ньюкасла, \\ Ньюкасл-апон-Тайн, Великобританія
\addr{label3} Вищий інститут комплексних професій, Тріполі, Лівія
}

\makeukrtitle

\begin{abstract}
Останнім часом представлено декілька досліджень стосовно наявності кисню в алмазі, наприклад теоретично і експериментально призначена модель
для N3 EPR створена з комплексу заміщувального азоту та заміщувального кисню як найближчих сусідів.
У цій статті представлено  ab initio обчислення заміщувального кисню в алмазі з огляду на стійкість, електронні структури, геометрію та надтонку взаємодію, а також   показано, що  заміщувальний кисень з C$_{2v}\,$, S = 1 є конфігурацією основного стану.
Встановлено, що кисень продукує або ж донорний, або акцепторний рівень в залежності від положення Фермі рівня.
\keywords теорія функціоналу густини, алмаз, кисень, надтонка взаємодія
\end{abstract}


\begin{thebibliography}{99}
\providecommand{\url}[1]{\texttt{#1}}
\providecommand{\urlprefix}{URL }
\providecommand{\eprint}[2][]{\url{#2}}

\bibitem{briddon-PB-185-179}
Briddon P.R., Jones R., Physica B, 1993, \textbf{185}, No. 1--4, 179; \bibdoi{10.1016/0921-4526(93)90235-X}.

\bibitem{jones-DRM-53-35}
Jones R., Goss J., Pinto H., Palmer D., Diamond Relat{.} Mater{.}, 2015,
  \textbf{53}, 35; \bibdoi{10.1016/j.diamond.2015.01.002}.

\bibitem{atumi-JPCM-25-6-065802}
Atumi M.K., Goss J.P., Briddon P.R., Shrif F.E., Rayson M.J., J. Phys.: Condens. Matter, 2013, \textbf{25}, No.~6, 065802; \bibdoi{10.1088/0953-8984/25/6/065802}.

\bibitem{chrenko-PRB.7.4560}
Chrenko R., Phys. Rev. B, 1973, \textbf{7}, 4560; \bibdoi{10.1103/PhysRevB.7.4560}.

\bibitem{farrer-SSC-7-685}
Farrer R., Solid State Commun{.}, 1969, \textbf{7}, 685; \bibdoi{10.1016/0038-1098(69)90593-6}.

\bibitem{crowther-PR-154-772}
Crowther P.A., Dean P.J., Sherman W.F., Phys{.} Rev{.}, 1967, \textbf{154},
  No.~3, 772;  \bibdoi{10.1103/PhysRev.154.772}.

\bibitem{prins-PRB-61-7191}
Prins J.F., Phys{.} Rev{.} B, 2000, \textbf{61}, No.~11, 7191;  \bibdoi{10.1103/PhysRevB.61.7191}.

\bibitem{Walker-N-263-275}
Walker J., Nature, 1976, \textbf{263}, No. 275, 275;  \bibdoi{10.1038/263275a0}.

\bibitem{melton-n-263-309}
Melton C.E., Nature, 1976, \textbf{263}, 309;  \bibdoi{10.1038/263309a0}.

\bibitem{ruan-APL-60-1379}
Ruan J., Choyke W.J., Kobashi K., Appl{.} Phys{.} Lett{.}, 1993, \textbf{60},
  No.~12, 1379; \bibdoi{10.1063/1.108685}.

\bibitem{mori-apl-60-47}
Mori Y., Eimori N., Kozuka H., Yokota Y., Moon H., Ma J.S., Ito T., Hiraki A.,
  Appl{.} Phys{.} Lett{.}, 1992, \textbf{60}, No.~1, 47;  \bibdoi{10.1063/1.107368}.

\bibitem{iakoubovskii-PRB-66-045406}
Iakoubovskii K., Stesmans A., Phys{.} Rev{.} B, 2002, \textbf{66}, No.~4,
  045406;  \bibdoi{10.1103/PhysRevB.66.045406}.

\bibitem{Ammerlaan1998} Ammerlaan C.A.J., In: Landolt-B\"ornstein Numerical Data and Functional Relationships in Science and Technology New Series Vol.~III/22b, Schultz~M. (Ed.), Springer, Berlin, 1989, 177--206.


\bibitem{etmimi-jpcm-22-385502}
Etmimi K.M., Goss J.P., Briddon P.R., Gsiea A.M., J{.} Phys{.}: Condens{.} Matter,
  2010, \textbf{22}, No.~38, 385502; \\ \bibdoi{10.1088/0953-8984/22/38/385502}.

\bibitem{komarovskikh-PSSA-210-2074}
Komarovskikh A., Nadolinny V., Palyanov Y., Kupriyanov I., Phys{.} Status
  Solidi A, 2013, \textbf{210}, No.~10, 2074; \\ \bibdoi{10.1002/pssa.201300036}.

\bibitem{palyanov-7-3169}
Palyanov Y.N., Borzdov Y.M., Khokhryakov A.F., Kupriyanov I.N., Sokol A.G.,
  Cryst. Growth Des., 2010, \textbf{10}, No.~7, 3169;  \bibdoi{10.1021/cg100322p}.

\bibitem{Komarovskikh-PSSA-31163}
Komarovskikh A., Nadolinny V., Palyanov Y., Kupriyanov I., Sokol A., Phys{.}
  Status Solidi A, 2014, \textbf{211}, No.~10, 2274;  \bibdoi{10.1002/pssa.201431163}.

\bibitem{gali-JPCM-13-11607}
Gali A., Lowther J.E., De{\'a}k P., J{.} Phys{.}: Condens{.} Matter, 2001,
  \textbf{13}, 11607;  \bibdoi{10.1088/0953-8984/13/50/319}.

\bibitem{perdew-prl-77-3865}
Perdew J.P., Burke K., Ernzerhof M., Phys{.} Rev{.} Lett{.}, 1996, \textbf{77},
  3865;  \bibdoi{10.1103/PhysRevLett.77.3865}.

\bibitem{briddon-pssb-217-131}
Briddon P.R., Jones R., Phys{.} Status Solidi B, 2000, \textbf{217}, No.~1,
  131; \\ \bibdoi{10.1002/(SICI)1521-3951(200001)217:1\%3C131::AID-PSSB131\%3E3.0.CO;2-M}.

\bibitem{rayson-cpc-178-128}
Rayson M.J., Briddon P.R., Comput. Phys{.} Commun{.}, 2008, \textbf{178}, No.~3,
  128;  \bibdoi{10.1016/j.cpc.2007.08.007}.

\bibitem{monkhorst-prb-13-5188}
Monkhorst H.J., Pack J.D., Phys{.} Rev{.} B, 1976, \textbf{13}, No.~12, 5188;  \bibdoi{10.1103/PhysRevB.13.5188}.

\bibitem{goss-tap-104-69}
Goss J.P., Shaw M.J., Briddon P.R., In:  Theory of Defects in Semiconductors,
Drabold~D.A., Estreicher~S.K. (Eds.), Springer, Berlin/Heidelberg, 2007, 69--94; \bibdoi{10.1007/11690320_4 }.

\bibitem{Sze1981}
Sze S.M., Physics of Semiconductor Devices, 2nd
  Edn., Wiley-Interscience, New York, 1981.

\bibitem{liberman-PRB-62-6851}
Liberman D.A., Phys{.} Rev{.} B, 2000, \textbf{62}, No.~11, 6851;  \bibdoi{10.1103/PhysRevB.62.6851}.

\bibitem{hartwigsen-prb-58-3641}
Hartwigsen C., Goedecker S., Hutter J., Phys{.} Rev{.} B, 1998, \textbf{58},
  No.~7, 3641;  \bibdoi{10.1103/PhysRevB.58.3641}.

\bibitem{shaw-prl-95-105502}
Shaw M.J., Briddon P.R., Goss J.P., Rayson M.J., Kerridge A., Harker A.H.,
  Stoneham A.M., Phys{.} Rev{.} Lett{.}, 2005, \textbf{95}, 205502;  \bibdoi{10.1103/PhysRevLett.95.205502}.

\bibitem{blochl-prb-50-17953}
Bl\"{o}chl P.E., Phys{.} Rev{.} B, 1994, \textbf{50}, No.~4, 17953;  \bibdoi{10.1103/PhysRevB.50.17953}.

\bibitem{rayson-pre-76-026704}
Rayson M.J., Phys{.} Rev{.} E, {2007}, \textbf{{76}}, No.~2,
  {026704};  \bibdoi{10.1103/PhysRevE.76.026704}.

\bibitem{watkins-PBC-117-9}
Watkins G.D., Physica B, 1983, \textbf{117--118}, No. 1--3, 9;  \bibdoi{10.1016/0378-4363(83)90432-1}.

\bibitem{coulson-PRSLA-241-433}
Coulson C.A., Kearsley M.J., Proc{.} R{.} Soc{.} London, Ser{.} A, 1957,
  \textbf{241}, 433;  \bibdoi{10.1098/rspa.1957.0138}.

\bibitem{dannefaer-DRM-10-2113}
Dannefaer S., Pu A., Kerr D., Diamond Relat{.} Mater{.}, 2001, \textbf{10},
  2113;  \bibdoi{10.1016/S0925-9635(01)00489-7}.

\end{thebibliography}
\end{document}